\documentclass[11pt,reqno]{amsart}
\usepackage[letterpaper, total={6.5in,9in}, includefoot, centering]{geometry}
\usepackage{amsfonts}
\usepackage{amsmath,amsthm,amssymb,amscd}
\usepackage[foot]{amsaddr}

\usepackage[mathscr]{eucal}
\usepackage{latexsym}
\usepackage{graphics}
\usepackage{verbatim}
\usepackage{upgreek}
\usepackage{overpic}
\usepackage[all,cmtip]{xy} 
\usepackage{enumitem}
\usepackage{tikz}
\usepackage{tikz-cd}
\usepackage[pagebackref,colorlinks=true,linkcolor=blue]{hyperref}
\renewcommand*\backref[1]{\ifx#1\relax \else (Page #1) \fi}
\usepackage[font=scriptsize]{caption}
\usepackage{upgreek}
\usepackage{cleveref}
\usepackage{float}

\usepackage[authoryear,sort&compress]{natbib}
\bibpunct{[}{]}{;}{n}{,}{,}

\hypersetup{
    colorlinks=true,
    linkcolor=blue,    
    citecolor=red,
    }

\makeatletter \@addtoreset{equation}{section}

\makeatother

\newtheorem{remark}{Remark}[section]

\newtheorem{prop}{Proposition}[section]

\graphicspath{ {./} }

\newcommand{\m}[1]{\mathcal{#1}}
\newcommand{\bb}[1]{\mathbb{#1}}

\begin{document}
\title{Galerkin Formulation of Path Integrals in Lattice Field Theory}
\author{Brian K. Tran$^{1,2,*}$ and Ben S. Southworth$^2$}
\address{$^1$ University of Colorado Boulder, Department of Applied Mathematics. Boulder, CO, USA 80309.}
\address{$^2$ Los Alamos National Laboratory, Theoretical Division. Los Alamos, NM, USA 87545.}
\address{$^*$ Corresponding author}
\email{brian.tran@colorado.edu, southworth@lanl.gov}
\allowdisplaybreaks

\begin{abstract}
We present a mathematical framework for Galerkin formulations of path integrals in lattice field theory. The framework is based on using the degrees of freedom associated to a Galerkin discretization as the fundamental lattice variables. We formulate standard concepts in lattice field theory, such as the partition function and correlation functions, in terms of the degrees of freedom. For example, using continuous finite element spaces, we show that the two-point spatial correlation function can be defined between any two points on the domain (as opposed to at just lattice sites) and furthermore, this two-point function satisfies a weak propagator (or Green's function) identity, in analogy to the continuum case, as well as a convergence estimate obtained from the standard finite element techniques. Furthermore, this framework leads naturally to higher-order formulations of lattice field theories by considering higher-order finite element spaces for the Galerkin discretization. We consider analytical and numerical examples of scalar field theory to investigate how increasing the order of piecewise polynomial finite element spaces affect the approximation of lattice observables. Finally, we sketch an outline of this Galerkin framework in the context of gauge field theories.
\end{abstract}

\maketitle

{\hypersetup{hidelinks} \tableofcontents}

\section{Introduction}\label{sec:intro}
We develop a mathematical framework for Galerkin discretizations of path integrals in lattice field theories. In our framework, we will see that the degrees of freedom associated to the Galerkin discretization play the fundamental role of the lattice field variables and allow for similar calculations to standard lattice field theories. In particular, we will show how the expectation value of observables can be expressed in terms of a basis of expectation values of polynomial functions of the degrees of freedom.

Galerkin discretizations of path integrals, and in particular, finite element discretizations of path integrals, have been considered previously in the literature. Finite element approximations of path integrals in lattice field theory were introduced in \cite{QFE0}; particularly, \cite{QFE0} introduces finite element approximations of path integrals for scalar fields and fermions using (piecewise) linear finite elements on flat spatial domains. This was later generalized to Riemannian manifolds in \cite{QFE1} again using linear finite elements. Examples of using linear finite elements for computations of path integrals in statistical field theories are further explored in \cite{HaGi23}. In a different direction, Galerkin approximations of the source partition function using global polynomial approximation spaces are considered in the source Galerkin method \cite{LaGu96, LaGu92v2, PeEmGu03}. From the perspective of the framework introduced here, these are all examples of Galerkin discretizations of path integrals. We will show how standard computations in lattice field theory can be performed with, for example, higher order polynomial finite element spaces; the key to this procedure is to formulate the discrete partition function and similarly, expectation values, in terms of degrees of freedom chosen for the Galerkin subspace. 

\textbf{Motivation.}
We provide several motivating reasons and advantages for considering Galerkin formulations of lattice field theories.

\emph{Convergence estimates}. By formulating lattice field theories using a Galerkin framework, the well-developed functional analytic theory of Galerkin error estimates (e.g., \cite{BrSc08}) can be applied to prove results for convergence of lattice quantities in the continuum limit. For example, we prove such a result for convergence of the two-point function for a free massive scalar field in \Cref{prop:two-point-convergence}. In particular, such convergence estimates may prove useful for establishing convergence of lattice field theories to their continuum limit.

\emph{Higher-order approximation spaces}. Higher-order approximation spaces can be readily incorporated into a Galerkin lattice field theory, which can yield reduced error and lattice artifacts compared to lower order spaces, even for the same number of global degrees of freedom. See \Cref{sec:analytical-example} and \Cref{sec:convergence}.

\emph{Weak versions of continuum identities}. Compared to a standard lattice transcription which approximates the differential operators appearing in the action, a Galerkin approximation instead approximates the action by restricting to a finite-dimensional subspace, while retaining the same differential operators, acting weakly on the finite-dimensional subspace. As a result, one can derive weak versions of continuum identities satisfied by the Galerkin lattice field theory. For example, in \Cref{sec:analytical-example}, we show that the Galerkin two-point function satisfies a weak propagator (or Green's function) identity.

\emph{Curved geometries}. Utilizing finite element spaces for the Galerkin subspace, it is both theoretically and algorithmically straightforward to define a Galerkin lattice field theory on curved geometries (e.g., for a discussion of lattice QCD in curved spacetime, see \cite{latticecurved}).

\emph{Symmetries and structure-preservation}. Many field theories exhibit symmetries and lattice transcriptions of such theories should retain those symmetries, as these symmetries encode fundamental conservation laws and constraints by Noether's first and second theorems, respectively. A well-known example is the $U(1)$-gauge symmetry of QED and the lattice Wilson action which preserves this symmetry \cite{wilsonprd}. On the other hand, there has been much recent and ongoing work in the Galerkin, particularly finite element, literature on the construction of symmetry-preserving (or structure-preserving) spaces. For example, the finite element exterior calculus \cite{feec1, feec2, feec3} framework, which preserves the aforementioned $U(1)$-gauge symmetry at the level of the algebra, as well as preserving the cohomology of the continuous theory; this could lead to promising methods for lattice theories with non-trivial topology \cite{LUSCHERnpb, nontrivialtop}. It was also shown in \cite{groupequivariant2, groupequivariant} how to construct group-equivariant interpolating spaces, particularly on symmetric spaces (e.g., the space of Lorentzian metrics). Thus, the Galerkin framework opens the possibility of incorporating recent advances in symmetry-preserving spaces to construct lattice field theories which retain their continuum symmetries.

\emph{Algorithmic considerations}. There are many well-developed finite element libraries, e.g., MFEM \cite{mfem, mfem-web}, FEniCS \cite{fenics1, fenics2, fenics3, fenics4}, and Firedrake \cite{FiredrakeUserManual}, which provide simple and efficient implementation of finite elements. To compute expectation values in lattice field theories using the Galerkin approach, one simply needs the degrees of freedom which come standard in most finite element libraries, as well as linear/nonlinear solvers, for which there are many highly efficient packages, e.g., \emph{hypre} \cite{falgout2002hypre}. As an alternative to using such solvers, it is also straightforward to incorporate Markov Chain Monte Carlo \cite{MCMC, latticeMC} methods into the Galerkin framework to compute expectation values, by replacing the standard lattice field variables with the degrees of freedom. We plan to explore this alternative direction in future work and will not discuss it further here.

\textbf{Organization.} This paper is organized as follows. In \Cref{sec:galerkin}, we define the Galerkin discretization of the partition function from two perspectives: viewing the discretization as arising from a finite-dimensional subspace approximation and equivalently, as arising from a set of degrees of freedom on the finite-dimensional subspace. The former perspective has a natural interpretation as a subspace approximation of the original partition function whereas the latter perspective is more convenient to work with computationally. We further show that these two perspectives are equivalent at the level of the variational principle in \Cref{sec:variational-principle}. In \Cref{sec:path-integrals}, we consider the Galerkin discretization of expectation values. We assume throughout a Euclidean metric signature and a coercivity estimate for the action so that these expectation values are well-defined. In this section, we first show that these Galerkin path integrals are well-defined assuming appropriate coercivity of the action. Subsequently, we construct a basis for expectation values of polynomial functions on the Galerkin subspace; as we will see, this basis consists of $n-$point correlation functions between degrees of freedom. We further show how $n-$point spatial correlation functions can be recovered from this basis; in particular, for continuous finite element spaces, the spatial correlation functions can be constructed between any spatial locations on the domain and one is not restricted to only nodal sites on a lattice. 

In \Cref{sec:analytical-example}, we consider as an analytical example the Galerkin discretization of path integrals for the free massive scalar field on an arbitrary spatial domain. We obtain the Galerkin two-point correlation function expressed in terms of finite element mass and stiffness matrices and show that it satisfies a weak analogue of the continuum propagator (or Green's function) identity. In \Cref{prop:two-point-convergence}, we derive a convergence result for the Galerkin two-point function using standard finite element techniques. In this section, we also consider higher order piecewise polynomial spaces on an infinite lattice and compare the momentum space propagator between different order piecewise polynomial spaces. We demonstrate that the higher order spaces produce better approximations of the momentum space propagator, even for the same number of degrees of freedom per unit length. By formulating path integrals in a general Galerkin sense, we are able to naturally apply the significant machinery developed in finite element methods, such as meshing complex domains. In \Cref{sec:convergence}, we additionally perform a convergence test to numerically verify the derived convergence estimate. Finally, in \Cref{sec:general-fem}, we discuss more general finite element spaces in the Galerkin framework, particularly sketching an outline of lattice gauge theories with higher-order finite element spaces.

\section{Galerkin Formulation of the Partition Function}\label{sec:galerkin}
We begin by providing a mathematical framework for the Galerkin discretization of an action functional and define a Galerkin partition function.

Let $S: X \rightarrow \mathbb{R}$ be an action functional on an infinite-dimensional Banach space $X$, $S: \phi \mapsto S[\phi]$. We consider the partition function associated to the action, formally
\begin{equation}\label{eq:path-integral-general}
    Z = \int_X D\phi \exp(-S[\phi]),
\end{equation}
and more generally, expectation values $\langle g \rangle = Z^{-1} \int_X D\phi\, g(\phi) \exp(-S[\phi])$. We will define a Galerkin formulation of the lattice prescription of such path integrals. For generality, we will first proceed abstractly but we will see subsequently how this prescription corresponds to a lattice discretization when $X$ is a function space over a domain and the Galerkin approximation is a finite element approximation.

\begin{remark}
    Throughout, we assume that the action $S$ is well-defined on the entire space $X$; this further ensures that its restriction to any finite-dimensional subspace, $\mathbb{S}_h$ defined below in \eqref{eq:discrete-action-def}, is well-defined. One may be interested in spaces $X$ for which the action is not well-defined at various limiting points in $X$, e.g., non-smooth fields. One possible approach to addressing such field theories would be to use a discontinuous Galerkin (DG) formulation \cite{arnold2002} which essentially defines a new action whose domain is a larger space containing non-smooth fields, but retains a consistency property, i.e., when the field is smooth, the new action agrees with the original action.
\end{remark}

Let $X_h$ be a finite-dimensional subspace of $X$, where $h>0$ is a discretization parameter indexing a family of finite-dimensional subspaces such that $\lim_{h \rightarrow 0^+} \dim(X_h) = \infty$. Let $\{f^i_h: X_h \rightarrow \mathbb{R}\}_{i=1}^{\dim(X_h)}$ be an associated unisolvent set of degrees of freedom (DOFs). That is, $\{f^i_h\}$ is a collection of linear functionals on $X_h$ such that the values $\{f^i_h(\phi_h)\}$ uniquely determine $\phi_h \in X_h$. Thus, the mapping defined by evaluating the degrees of freedom,
\begin{align}\label{eq:dof-evaluation}
    \bb{E}_h: X_h &\rightarrow \m{F}_h := \mathbb{R}^{\dim(X_h)} \\
        \phi_h &\mapsto \{f^i_h(\phi_h)\}_{i=1}^{\dim(X_h)}, \nonumber
\end{align}
is an isomorphism between vector spaces $X_h$ and $\m{F}_h$. We will now introduce the Galerkin approximation of a field theory, essentially by restricting the partition function and corresponding expectation values to the finite-dimensional subspace $X_h$. Note then that the Galerkin approximation of the field theory is completely specified by the choice of finite-dimensional subspace $X_h$. Once $X_h$ is chosen, the corresponding Galerkin approximation is independent of other choices; in particular, any unisolvent set of DOFs may be chosen, and by the above isomorphism \eqref{eq:dof-evaluation}, all such choices are equivalent; however, in practice, for a given $X_h$, a particular set of DOFs may be easier to compute.

To define a Galerkin approximation of the partition function, we will draw an analogy with quotienting or gauge fixing of the path integral measure in gauge field theory (see, e.g., \cite{PS95,Schwartz13, Srednicki07}). Let us suppose we have a projection onto the finite-dimensional subspace, $\bb{P}_h: X \rightarrow X_h$; the choice of such a projection will not change the definition of the Galerkin partition function and is used here simply for intuition of the definition. Furthermore, denoting the inclusion $i_h: X_h \hookrightarrow X$, we define the projection with codomain $X$ as $P_h = i_h \circ \bb{P}_h: X \rightarrow X$. Now, we define the following degenerate action on $X$,
\begin{align}\label{eq:degenerate-action}
    S_h: X &\rightarrow \mathbb{R} \\
       \phi &\mapsto S_h[\phi] := S[P_h\phi]. \nonumber
\end{align}
We would like to consider the partition function associated to $S_h$, $\int_X D\phi \exp(-S_h[\phi])$, as a candidate for a finite-dimensional approximation of the partition function. However, this expression will not work, since $S_h$ has an infinite-dimensional symmetry given by the kernel of the projection, $\m{G}:=\ker{P_h}$, acting additively on $X$; thus, we cannot make sense of the expression $\int_{X} D\phi \exp(-S_h[\phi])$ as it involves integration over orbits of a non-compact and furthermore infinite-dimensional symmetry group. This is similar in spirit to the issue which arises with gauge field theories \cite{PS95, Schwartz13, Srednicki07}. Analogous to gauge field theories, the resolution is to, in some manner, quotient the domain and the measure by the gauge symmetries, formally, $\int_{X/\m{G}} D\phi/\m{G} \exp(-S_h[\phi])$. Noting that the quotient $X / \m{G} \cong X_h$, we define the Galerkin approximation of the partition function as
\begin{equation}\label{eq:path-integral-galerkin}
    Z_h := \int_{X_h} D\phi_h \exp(-\bb{S}_h[\phi_h]),
\end{equation}
where the discrete action on $X_h$ is defined as
\begin{equation}\label{eq:discrete-action-def}
    \bb{S}_h := S_h \circ i_h = S \circ i_h : X_h \rightarrow \mathbb{R}.
\end{equation}
We make sense of the measure $D\phi_h := D\phi/\m{G}$ and integration over the (finite-dimensional) space $X_h$ appearing in \eqref{eq:path-integral-galerkin} by using the degrees of freedom associated to the Galerkin approximation. Letting $\phi_h^i:= f^i(\phi_h)$, we have that the values of the degrees of freedom $\vec{\phi}_h:=(\phi^1_h,\dots,\phi^{\dim(X_h)}_h)$ coordinatize $X_h$, where we use the notation $\vec{\phi}$ to denote vectors in $\m{F}_h$. Using this coordinatization, we define the measure $D\phi_h$ to be the pushforward measure of the standard measure $\prod_i d\phi^i_h$ on $\m{F}_h = \mathbb{R}^{\dim(X_h)}$ by $\bb{E}_h^{-1}$. By the change of variables theorem between the measure spaces $(X_h, D\phi_h)$ and $(\m{F}_h, \prod_i d\phi^i_h)$, the partition function can be expressed in terms of integration on $\m{F}_h$ as
\begin{equation}\label{eq:path-integral-galerkin-discrete}
    Z_h = \int_{X_h} D\phi_h \exp(-\bb{S}_h[\phi_h]) = \int_{\m{F}_h} \prod_{i=1}^{\dim(X_h)} d\phi^i_h \exp(-\mathfrak{S}_h[\vec{\phi}_h]),
\end{equation}
where we define the discrete action on the degrees of freedom as
$$ \mathfrak{S}_h = \bb{S}_h \circ \bb{E}_h^{-1} : \m{F}_h \rightarrow \mathbb{R}. $$

Thus, we have defined a Galerkin approximation of the partition function \eqref{eq:path-integral-general} from two perspectives. On one hand, the path integral is interpreted as an integral over the finite-dimensional function space $X_h$; this perspective is natural as it can be interpreted essentially as the restriction of the infinite-dimensional path integral to a finite-dimensional function space and will further be natural for deriving error bounds in numerical approximations of the path integral. On the other hand, using the degrees of freedom, the path integral can be expressed as an integral over $\mathbb{R}^{\dim(X_h)}$ which is useful in practice to analytically or numerically evaluate path integrals, as we will see in examples below. 

Of course, by construction, these two perspectives are equivalent. We will now show that these two perspectives are also equivalent in a classical sense, in that the classical states described by the variational principle are equivalent between the two perspectives. 

\subsection{The Variational Principle on Degrees of Freedom}\label{sec:variational-principle}
In the standard formulation of lattice field theories, field variables are transcribed to the lattice by specifying the values of the field at nodes (or more generally, on lattice sites described by $k-$dimensional faces corresponding to lattice discretizations of $k-$forms, see \Cref{sec:general-fem} for further discussion of such fields). From the Galerkin perspective, these correspond to particular choices of degrees of freedom, e.g., in the case of a scalar field the degrees of freedom correspond to nodal values of the field. Here, we show that the variational principle from the finite-dimensional function space perspective and the degrees of freedom perspective are equivalent, for the discrete actions $\mathbb{S}_h$ and $\mathfrak{S}_h$ defined above.

To summarize the definitions made above, we have the following commutative diagram: 
\begin{equation}\label{comm-diagram-def}
\begin{tikzcd}
	X \\
	\\
	{X_h} &&& {\mathbb{R}} \\
	\\
	{\mathcal{F}_h}
	\arrow["{S_h}", from=1-1, to=3-4]
	\arrow["{i_h}", from=3-1, to=1-1]
	\arrow["{\mathbb{S}_h}", from=3-1, to=3-4]
	\arrow["{\mathbb{E}_h}", shift left=2, from=3-1, to=5-1]
	\arrow["{\mathbb{E}_h^{-1}}", shift left=2, from=5-1, to=3-1]
	\arrow["{\mathfrak{S}_h}"', from=5-1, to=3-4]
\end{tikzcd}
\end{equation}
We have the following equivalence of the variational principle between the two perspectives.

\begin{prop}[Variational Principle on Degrees of Freedom]\label{Variational Principle on DOF}
The variational principle applied to the discrete action on $X_h$, $\bb{S}_h$, and applied to the discrete action on degrees of freedom, $\mathfrak{S}_h$, are equivalent. That is, the following diagram commutes:
\begin{equation}\label{comm-diagram-variational}
\begin{tikzcd}
	& {S: X\rightarrow\mathbb{R}} \\
	&& {} \\
	\\
	{\mathbb{S}_h:X_h\rightarrow\mathbb{R}} && {\mathfrak{S}_h: \mathcal{F}_h\rightarrow\mathbb{R}} \\
	\\
	\\
	& {\textup{DEL}}
	\arrow["{i_h^*}"', from=1-2, to=4-1]
	\arrow["{(i_h\circ \mathbb{E}_h^{-1})^*}", from=1-2, to=4-3]
	\arrow["{(\mathbb{E}_h^{-1})^*}", shift left=2, from=4-1, to=4-3]
	\arrow["{\textup{Variational Principle}}"', from=4-1, to=7-2]
	\arrow["{\mathbb{E}_h^*}", shift left=2, from=4-3, to=4-1]
	\arrow["{\textup{Variational Principle}}", from=4-3, to=7-2]
\end{tikzcd}.
\end{equation}
Here, \textup{DEL} denotes the discrete Euler--Lagrange equations arising from the variational principle and $i_h^*$ denotes pullback by $i_h$, i.e., $i_h^*S = S \circ i_h$.
\begin{proof}
The top loop of the diagram follows from the definitions, see diagram \eqref{comm-diagram-def}. The DEL arising from the variational principle for $\bb{S}_h$ is to find $\phi_h \in X_h$ such that $\delta \bb{S}_h[\phi_h]\cdot v_h = 0$ for all $v_h \in X_h$. On the other hand, the DEL for $\mathfrak{S}_h$ is to find $\vec{\phi}_h \in \mathcal{F}_h$ such that $\delta \mathfrak{S}_h[\vec{\phi}_h ]\cdot \vec{v}_h = 0$ for all $\vec{v}_h \in \mathcal{F}_h$. Then, one has
\begin{align*}
0 &= \delta \mathfrak{S}_d[\vec{\phi}_h]\cdot \vec{v}_h = \frac{d}{d\epsilon}\Big|_{\epsilon = 0}\mathfrak{S}_d[\vec{\phi}_h + \epsilon \vec{v}_h] = \frac{d}{d\epsilon}\Big|_{\epsilon = 0} \bb{S}_h[ \bb{E}_h^{-1}(\vec{\phi}_h + \epsilon \vec{v}_h)] \\
&= \frac{d}{d\epsilon}\Big|_{\epsilon = 0} \bb{S}_h[\bb{E}_h^{-1}\vec{\phi}_h + \epsilon \bb{E}_h^{-1}\vec{v}_h] = \delta \bb{S}_h[\bb{E}_h^{-1}\vec{\phi}_h]\cdot \bb{E}_h^{-1}\vec{v}_h.
\end{align*}
Since this must hold for all $\vec{v}_h \in \mathcal{F}_h$ and $\bb{E}_h^{-1}: \mathcal{F}_h \rightarrow X_h$ is an isomorphism, this is equivalent to finding $\vec{\phi}_h \in \mathcal{F}_h$ such that 
$$\delta \bb{S}_h[\bb{E}_h^{-1}\vec{\phi}_h]\cdot v_h = 0$$ 
for all $v_h \in X_h$. Again since $\bb{E}_h^{-1}$ is an isomorphism, finding such $\vec{\phi}_h \in \m{F}_h$ is equivalent to finding $\phi_h = \bb{E}_h^{-1}\vec{\phi}_h \in X_h$ such that $\delta \bb{S}_h[\phi_h]\cdot v_h =0$ for all $v_h \in X_h$. This is precisely the DEL for $\bb{S}_h$.
\end{proof}
\end{prop}

\subsection{Galerkin Formulation of Expectation Values}\label{sec:path-integrals}
Now that we have defined a Galerkin partition function \eqref{eq:path-integral-galerkin}, let us check that it is well-defined for fixed $h > 0$. To do so, it is generally necessary to make an assumption on the action $S$ so that the integral \eqref{eq:path-integral-galerkin} is convergent. A sufficient condition is to assume that $S$ is bounded from below, or coercive, in the following sense,
$$ S[\phi] \geq \alpha \|\phi\|_{X}^2 \text{ for all } \phi \in X, $$
for some constant $\alpha > 0$ 
\begin{remark}
This is the case, for example, for any Euclidean action of a field theory on a Wick-rotated spatial domain $D$ containing even degree terms with positive coefficients with quadratic coefficients bounded below by positive numbers, where $X=H^1(D)$. To see this, such an action can be expressed
\begin{align*}
    S[\phi] = \int_{D} \left[\frac{1}{2} a_0(x) \nabla\phi \cdot \nabla\phi + a_1(x) \phi^2 + \sum_{k=2}^{m}a_{k}(x)\phi^{2k}(x) \right] dx.
\end{align*}
Assuming $a_k(x) \geq 0$ for all $k$ and furthermore, $a_0(x) \geq \alpha_0 > 0$, $a_1(x) \geq \alpha_1 > 0$, then we have $S[\phi] \geq \alpha \|\phi\|^2_{H^1(D)}$ where $\alpha := \min\{\alpha_0,\alpha_1\}$.
\end{remark}
With such a coercivity assumption, noting that $\bb{S}_h = S \circ i_h$, we have
\begin{align*}
    Z_h &= \int_{X_h} D\phi_h \exp(-\bb{S}_h[\phi_h]) = \int_{\m{F}_h} \prod_{i=1}^{\dim(X_h)} d\phi^i_h \exp(-\mathfrak{S}_h[\vec{\phi}_h]) \\
    &= \int_{\m{F}_h} \prod_{i=1}^{\dim(X_h)} d\phi^i_h \exp(-\bb{S}_h(\bb{E}_h^{-1} \vec{\phi}_h )) \leq \int_{\m{F}_h} \prod_{i=1}^{\dim(X_h)} d\phi^i_h \exp(-\alpha \|\bb{E}_h^{-1} \vec{\phi}_h \|_X^2) \\
    &\leq \int_{\m{F}_h} \prod_{i=1}^{\dim(X_h)} d\phi^i_h \exp(- C(h)\alpha\|\vec{\phi}_h \|_{l^2}^2)
\end{align*}
where $C(h) > 0$ is a constant, for fixed $h>0$, arising from the equivalence of norms between isomorphic finite-dimensional normed vector spaces $(X_h, \|\cdot\|_X)$ and $(\m{F}_h,\|\cdot\|_{l^2})$ and $\|\cdot\|_{l^2}$ denotes the norm induced from the standard inner product on $\m{F}_h = \bb{R}^{\dim(X_h)}$. Finally, the right hand side is simply a finite-dimensional Gaussian integral, so $Z_h < \infty.$

Beyond the partition function, in lattice field theory, one is interested in expectation values, formally expressed as a path integral
$$ \langle g \rangle = Z^{-1}\int_X D\phi\, g(\phi)\exp(-S[\phi]), $$
where $\langle\cdot\rangle$ denotes the vacuum expectation value. In the Galerkin formulation, we analogously consider expectation values of functions $g_h: X_h \rightarrow \mathbb{R}$,
\begin{equation}\label{eq:general-galerkin-expectation}
    \langle g_h\rangle := Z_h^{-1}\int_{X_h} D\phi_h\, g_h(\phi_h) \exp(-\bb{S}_h[\phi_h]).
\end{equation}
Similar to the partition function, assuming at most modest growth on $g_h$, the expectation value \eqref{eq:general-galerkin-expectation} is well-defined assuming coercivity on the action. For example, this is the case if $g_h$ has polynomial growth, which will be our focus subsequently. Note that, by definition of $D\phi_h$ as the pushforward measure from the standard measure on $\m{F}_h$, \cref{eq:general-galerkin-expectation} is equivalently expressed in terms of degrees of freedom as
\begin{equation}\label{eq:general-dof-expectation}
    \langle g_h\rangle = Z_h^{-1}\int_{\m{F}_h} \prod_{i=1}^{\dim(X_h)} d\phi^i_h\, g_h \circ \bb{E}_h^{-1} (\vec{\phi}_h) \exp(-\mathfrak{S}_h[\vec{\phi}_h]).
\end{equation}
In particular, often one considers expectation values where $g_h$ is a polynomial in $\phi_h$. We will now construct a basis for such polynomial expectation values in terms of the degrees of freedom. Consider an $n-$multilinear form on $X_h$,
$$ M_h: \underbrace{X_h \times \dots \times X_h}_{n \text{ times}} \rightarrow \mathbb{R}, $$
satisfying 
\begin{equation}\label{eq:monomial-identity}
     M_h\left(\sum_{i_1} c^1_{i_1} \boldsymbol{\chi}^1_{i_1}, \dots, \sum_{i_n} c^n_{i_n} \boldsymbol{\chi}^n_{i_n}\right) = \sum_{i_1,\dots,i_n}c^1_{i_1}\cdots c^n_{i_n} M_h(\boldsymbol{\chi}^1_{i_1},\dots,\boldsymbol{\chi}^n_{i_n}) 
\end{equation}
for all $c^a_j \in \mathbb{R}$ and $\boldsymbol{\chi}^a_j \in X_h$. Since polynomials can be built as a linear combination of monomials, we consider the expectation value of a monomial function of degree $n$, $m_h(\phi_h) := M_h(\phi_h,\dots,\phi_h)$. From \eqref{eq:general-dof-expectation}, we have
\begin{equation*}
    \langle m_h\rangle = Z_h^{-1}\int_{\m{F}_h} \prod_{i=1}^{\dim(X_h)} d\phi^i_h M_h(\bb{E}_h^{-1} \vec{\phi}_h,\dots,\bb{E}_h^{-1} \vec{\phi}_h) \exp(-\mathfrak{S}_h[\vec{\phi}_h]).
\end{equation*}
Let $\{\mathbf{v}_i\}$ be a basis satisfying $\bb{E}_h \mathbf{v}_i = \vec{e}_i$ where $\vec{e}_i$ is the $i^{th}$ standard unit vector. Thus, $\bb{E}_h^{-1}\phi_h = \phi^i_h \mathbf{v}_i$ (using summation convention where repeating upper and lower indices are summed over) and using \cref{eq:monomial-identity}, the above expression can be expressed as
\begin{align*}
    \langle m_h\rangle &= Z_h^{-1}\int_{\m{F}_h} \prod_{i=1}^{\dim(X_h)} d\phi^i_h M_h(\bb{E}_h^{-1} \vec{\phi}_h,\dots,\bb{E}_h^{-1} \vec{\phi}_h) \exp(-\mathfrak{S}_h[\vec{\phi}_h]) \\
    &= Z_h^{-1}\int_{\m{F}_h} \prod_{i=1}^{\dim(X_h)} d\phi^i_h M_h(\phi^{i_1}_h \mathbf{v}_{i_1},\dots,\phi^{i_n}_h \mathbf{v}_{i_n}) \exp(-\mathfrak{S}_h[\vec{\phi}_h])  \\
    &= Z_h^{-1}\int_{\m{F}_h} \prod_{i=1}^{\dim(X_h)} d\phi^i_h \phi^{i_1}_h \cdots \phi^{i_n}_h M_h(\mathbf{v}_{i_1},\dots,\mathbf{v}_{i_n}) \exp(-\mathfrak{S}_h[\vec{\phi}_h]) \\
    &= M_h(\mathbf{v}_{i_1},\dots,\mathbf{v}_{i_n}) \left( Z_h^{-1}\int_{\m{F}_h} \prod_{i=1}^{\dim(X_h)} d\phi^i_h \phi^{i_1}_h \cdots \phi^{i_n}_h \exp(-\mathfrak{S}_h[\vec{\phi}_h]) \right).
\end{align*}
Thus, the expectation value of any monomial of degree $n$ can be expressed as a linear combination of the quantity appearing in the above parentheses.  This gives have a basis for expectation values of monomials of degree $n$,
\begin{equation}\label{eq:correlation-dofs}
    \langle \phi^{i_1}_h \cdots \phi^{i_n}_h\rangle := Z_h^{-1}\int_{\m{F}_h} \prod_{i=1}^{\dim(X_h)} d\phi^i_h \phi^{i_1}_h \cdots \phi^{i_n}_h \exp(-\mathfrak{S}_h[\vec{\phi}_h]).
\end{equation}
We refer to \eqref{eq:correlation-dofs} as the $n-$point correlation function on degrees of freedom. Analogous to $n-$point spatial correlation functions of a field theory, these correlation functions on degrees of freedom carry all of the information of the Galerkin lattice field theory, as they can be used to compute the expectation value of any polynomial function on $X_h$. Rather abstractly, \eqref{eq:correlation-dofs} gives the correlation between the $i_1^{th}, \dots, i_n^{th}$ degrees of freedom. To see this interpretation more clearly, we can transform \eqref{eq:correlation-dofs} back into integration over $X_h$,
\begin{equation*}
        \langle \phi^{i_1}_h \cdots \phi^{i_n}_h\rangle = Z_h^{-1}\int_{X_h} D\phi_h f^{i_1}(\phi_h) \cdots f^{i_n}(\phi_h) \exp(-\bb{S}_h[\phi_h]).
\end{equation*}
It is clear that when $f^i$ are chosen to be the evaluation linear functionals $f^i(\phi_h) = \phi_h(x_i)$ (e.g., corresponding to a nodal finite element space), the above expression reduces to the $n-$point spatial correlation function between nodes $x_{i_1},\dots,x_{i_n}$. However, let us now show that we can more generally recover the $n-$point spatial correlation functions for any choice of degrees of freedom, when $X$ is some suitable function space on a spatial domain $D$ and the Galerkin subspace $X_h$ is continuous, i.e., $X_h \subset C^0(D)$, such as is the case with standard conforming polynomial finite element methods. Furthermore, we will see that we can define the spatial correlation function for any points $x_1,\dots,x_n \in D$ and are not restricted to nodes on a lattice.

Note that the Galerkin formulation is not restricted to scalar fields, so we allow the possibility of color indices, which we denote $[\phi_h]^a$ and similarly for the basis functions defined above $[v_i]^a$. We consider the $n-$point spatial correlation function, for $\phi_h \in X_h$ and any $x_1,\dots,x_n \in D$,
$$ \langle [\phi_h]^{a_1}(x_1) \cdots [\phi_h]^{a_n}(x_n) \rangle, $$
which is defined since $X_h \subset C^0(D)$. Then, from the above calculation, we have that the $n-$point spatial correlation function can be expressed in terms of the $n-$point correlation functions on degrees of freedom as
$$ \langle [\phi_h]^{a_1}(x_1) \cdots [\phi_h]^{a_n}(x_n) \rangle = [\mathbf{v}_{i_1}]^{a_1}(x_1) \cdots [\mathbf{v}_{i_n}]^{a_n} (x_n) \langle \phi^{i_1}_h \cdots \phi^{i_n}_h\rangle.  $$
Interestingly, by using the Galerkin construction with a $C^0$ space, one can define the $n-$point spatial correlation function between any points $x_1,\dots,x_n \in D$ and one is not limited to just correlation functions between points on a lattice. In \Cref{sec:analytical-example}, we contextualize this abstraction by considering a particular example of a free massive scalar field using a finite element discretization and show that the two-point spatial correlation function can be expressed in terms of the finite element stiffness and mass matrices.

To conclude this section, using a standard statistical field theory argument, we note that the correlation functions on degrees of freedom are generated by a generating functional. Define the source partition function on degrees of freedom
\begin{equation}\label{eq:source-galerkin-partition-function}
    Z_h[\vec{J}] := \int_{\m{F}_h} \prod_{i=1}^{\dim(X_h)} d\phi^i_h \exp(-\mathfrak{S}_h[\vec{\phi}_h] + \vec{J}\cdot \vec{\phi}_h)
\end{equation}
This is again well-defined assuming coercivity of the action. Then, the source partition function is a generating functional for the $n-$point correlation function on degrees of freedom in the following sense,
\begin{equation}\label{eq:source-galerkin-partition-derivative}
    \langle \phi^{i_1}_h \cdots \phi^{i_n}_h\rangle = Z_h[\vec{J}]^{-1} \frac{\partial^n Z_h[\vec{J}]}{\partial J^{i_1} \cdots \partial J^{i_n}}\Big|_{\vec{J} = 0}.
\end{equation}

We will use this generating functional in the example of \Cref{sec:analytical-example}, which we previously alluded to, to compute the two-point spatial correlation function.

\subsubsection{Analytical Example: Free Massive Scalar Field}\label{sec:analytical-example}
\sloppy As an analytical example of the Galerkin formulation of path integrals, we will compute the partition function and propagator for a free massive scalar field using a finite element discretization in two cases. First, we will consider a domain in $\mathbb{R}^n$ and a general finite element discretization. Subsequently, we will consider an infinite regular lattice with piecewise-polynomial elements in one-dimension to explicitly compute the momentum space propagator and compare the result to the usual lattice discretization of a free massive scalar field. 

\textbf{Domain in $\mathbb{R}^n$.} Consider the (Euclidean) action for a massive scalar field,
$$S[\phi] = \int_D \left[\frac{1}{2} \lambda \nabla \phi \cdot \nabla \phi + \frac{1}{2}m^2\phi^2 \right] dx,$$
where $\lambda$ and $m^2$ are positive real-valued functions on a spatial domain $D$ in $\mathbb{R}^n$ with prescribed boundary conditions (either periodic, homogeneous Dirichlet, or homogeneous Neumann). Let $X_h$ be a finite-dimensional subspace of the space of $H^1$ functions on $D$ satisfying the prescribed boundary conditions, $X = H^1_{\text{BC}}(D)$. Expand $\phi_h \in X_h$ as $\phi_h = \phi^i_h \mathbf{v}_i$ where $\{\phi^i\}$ are the degrees of freedom with associated basis $\{\mathbf{v}_i\}$ of $X_h$. The discrete action on degrees of freedom is given by
\begin{align*}
\mathfrak{S}_h[\vec{\phi}_h] &= \int_D \left[\frac{1}{2} \lambda \nabla(\phi^i_h\mathbf{v}_i)\cdot\nabla(\phi^j_h \mathbf{v}_j) + \frac{1}{2}m^2 (\phi^i_h\mathbf{v}_i)( \phi^j_h\mathbf{v}_j) \right] dx \\
&= \phi^i_h \left(\int_D \left[ \frac{1}{2} \lambda \nabla \mathbf{v}_i \cdot \nabla \mathbf{v}_j + \frac{1}{2}m^2 \mathbf{v}_i\mathbf{v}_j \right] dx \right) \phi^j_h \\
&= \frac{1}{2}\vec{\phi}^T_h (K+M) \vec{\phi}_h,
\end{align*}
where we introduced the stiffness matrix with components $K_{ij} := \int_D \lambda \nabla \mathbf{v}_i \cdot \nabla \mathbf{v}_j dx$ and the mass matrix with components $M_{ij} := \int_D m^2 \mathbf{v}_i\mathbf{v}_j dx.$ Note that $K+M$ is a symmetric positive-definite matrix and hence, invertible. Since $\mathfrak{S}_h$ is quadratic in the degrees of freedom, the associated partition function can be computed explicitly as a Gaussian integral
\begin{equation}\label{eq:fem-partition-function}
    Z_h = \prod_i \int_{-\infty}^{\infty} d\phi^i_h \exp(-\mathfrak{S}_h[\vec{\phi}_h]) = \det(2\pi (K+M)^{-1})^{1/2}.
\end{equation}
Similarly, the source partition function on degrees of freedom \eqref{eq:source-galerkin-partition-function} is given by
\begin{align}\label{eq:fem-source-partition-function}
    Z_h[\vec{J}] &= \prod_i \int_{-\infty}^{\infty} d\phi^i_h \exp(-\mathfrak{S}_h[\vec{\phi}_h] + \vec{J}\cdot \vec{\phi}_h) 
    \\ &= \det(2\pi (K+M)^{-1})^{1/2} \exp\left(\frac{1}{2} \vec{J}^T (K+M)^{-1} \vec{J}\right). \nonumber
\end{align}
By differentiating the source partition function on degrees of freedom, one can compute the $n-$point function on degrees of freedom, \eqref{eq:source-galerkin-partition-derivative}. For example, for the case $n=2$, we have
\begin{equation}\label{eq:fem-two-point-function}
    \langle \phi^i_h\phi^j_h \rangle = Z^{-1}_h \prod_k \int_{-\infty}^{\infty} d\phi^k_h \phi^i_h \phi^j_h \exp(-\mathbb{S}_d[\vec{\phi}_h]) = Z_h[\vec{J}]^{-1} \frac{\partial^2 Z_h[\vec{J}]}{\partial J^i \partial J^j}\Big|_{\vec{J}=0} = (K+M)^{-1}_{ij}.
\end{equation}
Note that, unless the finite element space is a nodal finite element space with degrees of freedom given by evaluation at the nodes, \cref{eq:fem-two-point-function} is not the two-point function between nodes $i$ and $j$ but is instead, rather abstractly, the two-point function between the $i^{th}$ and $j^{th}$ degree of freedom. However, as discussed in \Cref{sec:path-integrals}, an advantage of the Galerkin formulation with a $C^0$ finite element space is that we can compute the two-point function between any two positions and are not restricted to only nodes on a lattice. Assuming a $C^0$ finite element space, consider the two-point function, for $x,y \in D$, 
$$ \langle \phi_h(x)\phi_h(y)\rangle. $$
Expanding in the basis $\{\mathbf{v}_i\}$, we obtain an explicit expression for the two-point function in terms of the mass and stiffness matrices,
\begin{align}\label{eq:two-point-function-analytic}
    \langle \phi_h(x)\phi_h(y)\rangle &= \langle \phi^i_h \mathbf{v}_i(x) \phi^j_h \mathbf{v}_j(y) \rangle = \mathbf{v}_i(x) \langle \phi^i_h \phi^j_h\rangle \mathbf{v}_j(y) = \vec{\mathbf{v}}(x)^T (K+M)^{-1} \vec{\mathbf{v}}(y).
\end{align}
Note that the expression \cref{eq:two-point-function-analytic} is rather general; it provides a Galerkin approximation of the two-point function for any $C^0$ finite element approximation on a domain $D$ in $\mathbb{R}^n$, for any $x,y \in D$. Furthermore, the Galerkin two-point function \cref{eq:two-point-function-analytic} satisfies a discrete analogue of the continuum propagator (or Green's function) identity
$$ (-\nabla_x\cdot (\lambda(x) \nabla_x) + m^2(x))\langle \phi(x)\phi(y)\rangle = \delta(x-y). $$
In particular, for any $\psi \in X_h$, we can expand $\psi(x) = \psi^i \mathbf{v}_i(x)$. Then, using \eqref{eq:two-point-function-analytic},
\begin{align*}
    \int_D \psi(x)  (-\nabla_x\cdot & (\lambda(x) \nabla _x) + m^2(x))\langle \phi_h(x)\phi_h(y)\rangle dx \\
    &= \int_D \psi^i \mathbf{v}_i(x) (-\nabla_x\cdot (\lambda(x) \nabla_x) + m^2(x)) \sum_{jk} \mathbf{v}_j(x) (K+M)^{-1}_{jk} \mathbf{v}_k(y) dx \\
    &= \psi^i \sum_{jk} \int_D(\lambda(x) \nabla_x \mathbf{v}_i(x) \cdot \nabla_x \mathbf{v}_j(x) + m^2(x) \mathbf{v}_i(x) \mathbf{v}_j(x)) dx (K+M)^{-1}_{jk} \mathbf{v}_k(y) \\
    &= \psi^i \sum_{jk} (K+M)_{ij} (K+M)^{-1}_{jk} \mathbf{v}_k(y) \\
    &= \psi^i \sum_k \delta_{ik} \mathbf{v}_k(y) = \psi^i \mathbf{v}_i(y) = \psi(y). 
\end{align*}
Interestingly, this says that the Galerkin two-point function $\langle \phi_h(x)\phi_h(y)\rangle$ satisfies the propagator identity weakly on functions in $X_h$. 

For the case of a free massive scalar field, it is straight-forward to obtain a convergence result for the two-point propagator, arising from using the expression \eqref{eq:fem-two-point-function} of the two-point function in terms of the finite element mass and stiffness matrices, paired with standard finite element convergence estimates.

\begin{prop}\label{prop:two-point-convergence}
    Suppose further that the functions $\lambda$ and $m^2$ are bounded from below by a positive constant and bounded from above. Let $f \in L^2(D)$ and denote by $\vec{f}$ the vector with components
    $$ f_i := (\mathbf{v}_i, f)_{L^2(D)},\ i=1,\dots,\dim(X_h). $$
    Then, 
    $$ \Big\|\sum_{ij} \mathbf{v}_i \langle \phi^i_h \phi^j_h\rangle f_i - (-\nabla\cdot (\lambda \nabla) + m^2)^{-1} f \Big\|_{H^1(D)} \leq C \min_{u^* \in X_h} \Big\|u^* - (-\nabla\cdot (\lambda \nabla) + m^2)^{-1} f \Big\|_{H^1(D)}, $$
    where $C$ depends on (the bounds of) $\lambda$ and $m^2$ but is independent of $f$.
    \begin{proof}
        Using the identity relating the two-point function on degrees of freedom to the finite element mass and stiffness matrices \eqref{eq:fem-two-point-function}, the quantity 
        $$u_h := \sum_{ij} \mathbf{v}_i \langle \phi^i_h \phi^j_h\rangle f_i \in X_h $$
        is equivalently the solution of the variational problem
        $$ a(u_h, v_h) = (v_h, f)_{L^2(D)} \text{ for all } v_h \in X_h, $$
        where the bilinear form $a$ is given by
        $$ a(u,v) = \int_D (\lambda \nabla u \cdot \nabla v + m^2 uv) dx. $$
        Similarly, the quantity $u := (-\nabla\cdot (\lambda \nabla) + m^2)^{-1} f \in H^1_{\text{BC}}(D)$ is equivalently given by the solution of the variational problem 
        $$ a(u, v) = (v, f)_{L^2(D)} \text{ for all } v \in H^1_{\text{BC}}(D). $$
        The result now immediately follows from C\'{e}a's lemma (see, e.g., \cite{BrSc08}).
    \end{proof}
\end{prop}

In particular, for a standard $H^1$-conforming piecewise polynomial finite element space of degree $p$ on a regular triangulation, we expect degree $p$ convergence in $H^1(D)$ (see \cite{BrSc08}). We demonstrate this numerically in \Cref{sec:convergence} by considering the two-point function for the scalar field on a conformal two-sphere. 

\textbf{One-dimensional infinite lattice.} It is instructive to consider a concrete example; consider the case of a regular infinite lattice $\mathbb{T}$ on $\mathbb{R}$ with nodes $\{x_j = jh\}$ and spacing $h > 0$ in one dimension, using piecewise-linear elements satisfying $\mathbf{v}_i(x_j) = \delta_{ij}$. In this case, a unisolvent set of degrees of freedom are given by the evaluation functionals $f^i_h(\phi_h) := \phi_h(x_i)$. Thus, the degrees of freedom are just the nodal values of the discrete field. The stiffness and mass matrices are given by
$$ (K+M)_{ij} = \underbrace{h^{-1}(-\delta_{i+1,j} + 2\delta_{i,j} - \delta_{i-1,j})}_{= \int \nabla \mathbf{v}_i \cdot\nabla \mathbf{v}_j dx} + \underbrace{h m^2 \left(\frac{1}{6} \delta_{i+1,j} + \frac{2}{3}\delta_{i,j} + \frac{1}{6}\delta_{i-1,j}\right)}_{= m^2 \int \mathbf{v}_i \mathbf{v}_j dx}. $$

In momentum space, a direct calculation via the Fourier transform yields an expression for the reciprocal of the momentum space propagator,
\begin{equation}\label{eq:one-d-momentum-propagator-fem}
    \langle \phi_h(-p)\phi_h(p)\rangle^{-1} = 4h^{-2} \sin(p h/2)^2 + m^2\left(\frac{2}{3} + \frac{1}{3}\cos(ph) \right).
\end{equation}

On the other hand, in the standard formulation of a lattice scalar field \cite{Smit02,He2012,Creutz23}, the mass of $\phi^i$ is entirely localized at the node $i$ (whereas in the case of using piecewise-linear elements, $\phi^i$ contributes mass to nodes $i-1$ and $i+1$). The stiffness and mass matrices in the standard formulation are then given by
$$ (K'+M')_{ij} = h^{-1}(-\delta_{i+1,j} + 2\delta_{i,j} - \delta_{i-1,j}) + h m^2\delta_{ij}, $$
where we use $'$ to denote quantities associated to the standard lattice formulation. The associated reciprocal of the momentum space propagator is
\begin{equation}\label{eq:one-d-momentum-propagator-standard}
    \langle \phi'_h(-p)\phi'_h(p)\rangle^{-1} = 4h^{-2} \sin(p h/2)^2 + m^2.
\end{equation}
Expanding the two expressions \eqref{eq:one-d-momentum-propagator-fem} and \eqref{eq:one-d-momentum-propagator-standard} in $h$, we see that
\begin{align*}
\langle \phi_h(-p)\phi_h(p)\rangle^{-1} &= p^2 + m^2 + \mathcal{O}(h^2),\\
\langle \phi'_h(-p)\phi'_h(p)\rangle^{-1} &= p^2 + m^2 + \mathcal{O}(h^2).
\end{align*}
Thus, the two differing transcriptions produce the same continuum result to $\mathcal{O}(h^2)$. In finite element theory, this process of localizing the mass to a node is known as mass lumping; in this case, the process was to replace the tridiagonal matrix $M$ by a diagonal matrix $M'$ such that the resulting discretization is still consistent to the same order. This provides an elementary example of how techniques in finite element theory carry over to the transcription of lattice fields.

\begin{figure}[H]
\begin{center}
\includegraphics[width=110mm]{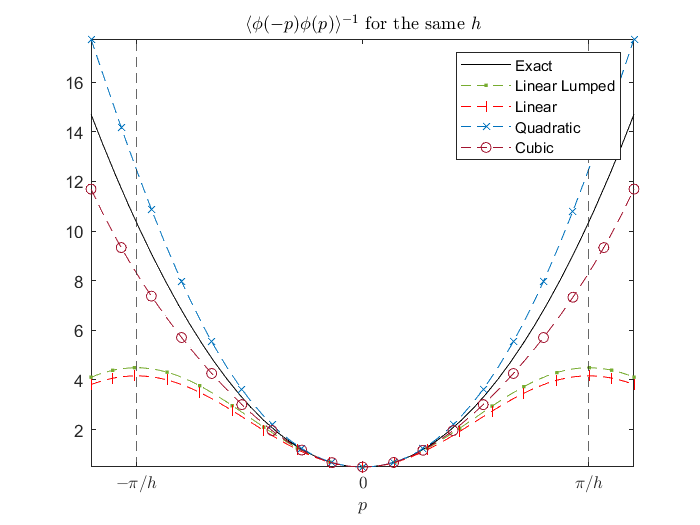}
\caption{Comparison of momentum space propagator with piecewise linear, piecewise linear lumped, piecewise quadratic and piecewise cubic elements for the same $h$ with $h = 1, m = 0.7$. Dashed lines indicate the Brillouin zone $[-\pi/h,\pi/h]$.} 
\label{figure:prop-same-h}
\end{center}
\end{figure}

\begin{figure}[H]
\begin{center}
\includegraphics[width=110mm]{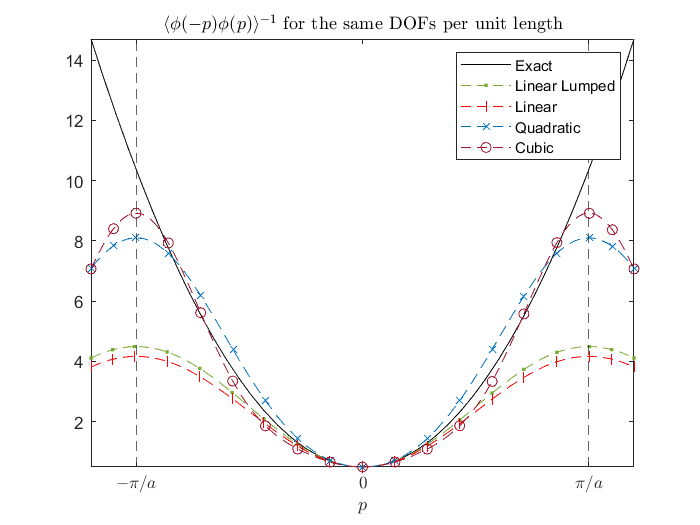}
\caption{Comparison of momentum space propagator with piecewise linear, piecewise linear lumped, piecewise quadratic and piecewise cubic elements for the same number of degrees of freedom per unit length with $a = 1, m = 0.7$. Dashed lines indicate the Brillouin zone $[-\pi/a,\pi/a]$.} 
\label{figure:prop-same-dof}
\end{center}
\end{figure}

In the same vein, we consider a finite element analogy to the Symanzik improvement program for constructing higher order approximations of the action; in the Symanzik improvement program, this is achieved by adding counterterms to the action to produce approximations higher than linear order in the lattice spacing \cite{Improvement1, Improvement2}. Analogously, we consider enriching the finite element space by considering higher order piecewise-polynomial spaces which in turns provides a higher order approximation of the action (see, e.g., \cite{Var1,Var2}). Analogous to the piecewise linear case, we compute the momentum space propagator, now using piecewise quadratic elements with an additional degree of freedom placed halfway between consecutive nodes $x_j, x_{j+1}$ and for piecewise cubic elements with two additional degrees of freedom equally spaced between consecutive nodes $x_j, x_{j+1}$. This is shown in \Cref{figure:prop-same-h}.
A better comparison in terms of computational complexity is to compare the momentum space propagator from the different finite element spaces using the same number of degrees of freedom per unit length. For a lattice spacing $a > 0$, we compare piecewise linear elements with $h=a$, piecewise quadratic elements with $h = 2a$, and piecewise cubic elements with $h=3a$. This is shown in \Cref{figure:prop-same-dof}.
Interestingly, even for the same number of degrees of freedom per unit length, the propagator obtained from the quadratic and cubic elements are better approximations of the exact propagator, particularly near the boundary of the Brillouin zone. In finite element terminology, this is a comparison of ``$h-$refinement" versus ``$p-$refinement" (see, e.g., \cite{BrSc08}); to produce a finer discretization, one can make $h$ smaller or alternatively, increase the polynomial degree of the finite element space. \Cref{figure:prop-same-dof} shows that improving the polynomial degree even while lowering the grid resolution, such that the number of degrees of freedom per unit length is constant, can lead to improvement in the approximation of the propagator. The improvement is particularly drastic near the 
boundary of the Brillouin zone, which could be explained by the fact that the higher degree polynomial spaces can better capture oscillatory behavior in the high energy Brillouin zone compared to piecewise linear elements. 

\subsubsection{Convergence Test}\label{sec:convergence}
As a numerical example, we consider a scalar field on the two-sphere equipped with a conformal metric,
$$S[\phi] = \int_{\bb{S}^2} \left[\frac{1}{2} \lambda \nabla \phi \cdot \nabla \phi + \frac{1}{2}\phi^2 \right] ds,$$
where $\lambda$ is the conformal factor corresponding to a conformal metric relative to the standard metric $g$ on $\bb{S}^2$; above, $ds$ denotes the surface measure on $\mathbb{S}^2$ relative to the standard metric. We take the conformal factor $\lambda$ to be 
$$ \lambda(x,y,z) = 1 + \exp(-2x^2 - 3y^2 - 2z^2), $$
expressed in the embedded coordinates $\bb{S}^2 \hookrightarrow \bb{R}^3$. The finite element formulation easily allows incorporation of such spatially varying coefficients (we could have also included such a spatially varying coefficient for the potential term $\phi^2$), which arises, for example, in lattice field theories on curved spacetime where the curvature is introduced through the metric \cite{latticecurved}. Since $\lambda$ is positive and bounded both above and below on $\bb{S}^2$, the relevant function space to consider is $X = H^1(\bb{S}^2)$.

From \Cref{sec:analytical-example}, the two-point propagator using a finite element discretization of $H^1(\bb{S}^2)$ is
$$     \langle \phi_h(x)\phi_h(y)\rangle = \vec{\mathbf{v}}(x)^T (K+M)^{-1} \vec{\mathbf{v}}(y),$$
where the stiffness and mass matrices are
$$ K_{ij} := \int_{\bb{S}^2} \lambda \nabla \mathbf{v}_i \cdot \nabla \mathbf{v}_j dS,\quad M_{ij} := \int_{\bb{S}^2} \mathbf{v}_i\mathbf{v}_j ds. $$
\begin{remark}
    The finite element discretization for this test was performed using the finite element library MFEM \cite{mfem, mfem-web}. The basis functions $\mathbf{v}_i$ are Lagrange polynomials which interpolate the Gauss-Legendre points in a given element, i.e., they satisfy $f^j(\mathbf{v}_i) = \delta_{ij}$, where the degrees of freedom $f^j$ are the pointwise evaluation functionals at the Gauss-Legendre points. 
\end{remark}
Thus, we can compute the two-point function between two points $x,y \in \bb{S}^2$ by solving the linear system
\begin{equation}\label{eq:Ax=b}
    (K+M)\vec{z} = \vec{\mathbf{v}}(y)
\end{equation}
and subsequently, computing the product $\vec{\mathbf{v}}(x)^T \vec{z}$. For finite element weak forms (and more generally sparse discretization of differential operators), there is significant research and software libraries for the fast parallel solution of \eqref{eq:Ax=b}. Multilevel methods, notably geometric and algebraic multigrid \cite{briggs2000multigrid}, can typically provide the solution in $O(N)$ FLOPs and scale excellent in parallel. Such methods have been extended to certain types of lattice field theory with great success \cite{multigridMC3,brannick2008adaptive,multigridMC1}. By utilizing FEM technology and software packages like MFEM, we can easily also use multilevel solver technology and packages such as \emph{hypre} \cite{falgout2002hypre} to rapidly compute two-point correlations. 

\begin{figure}[h!]
\begin{center}
\includegraphics[width=110mm]{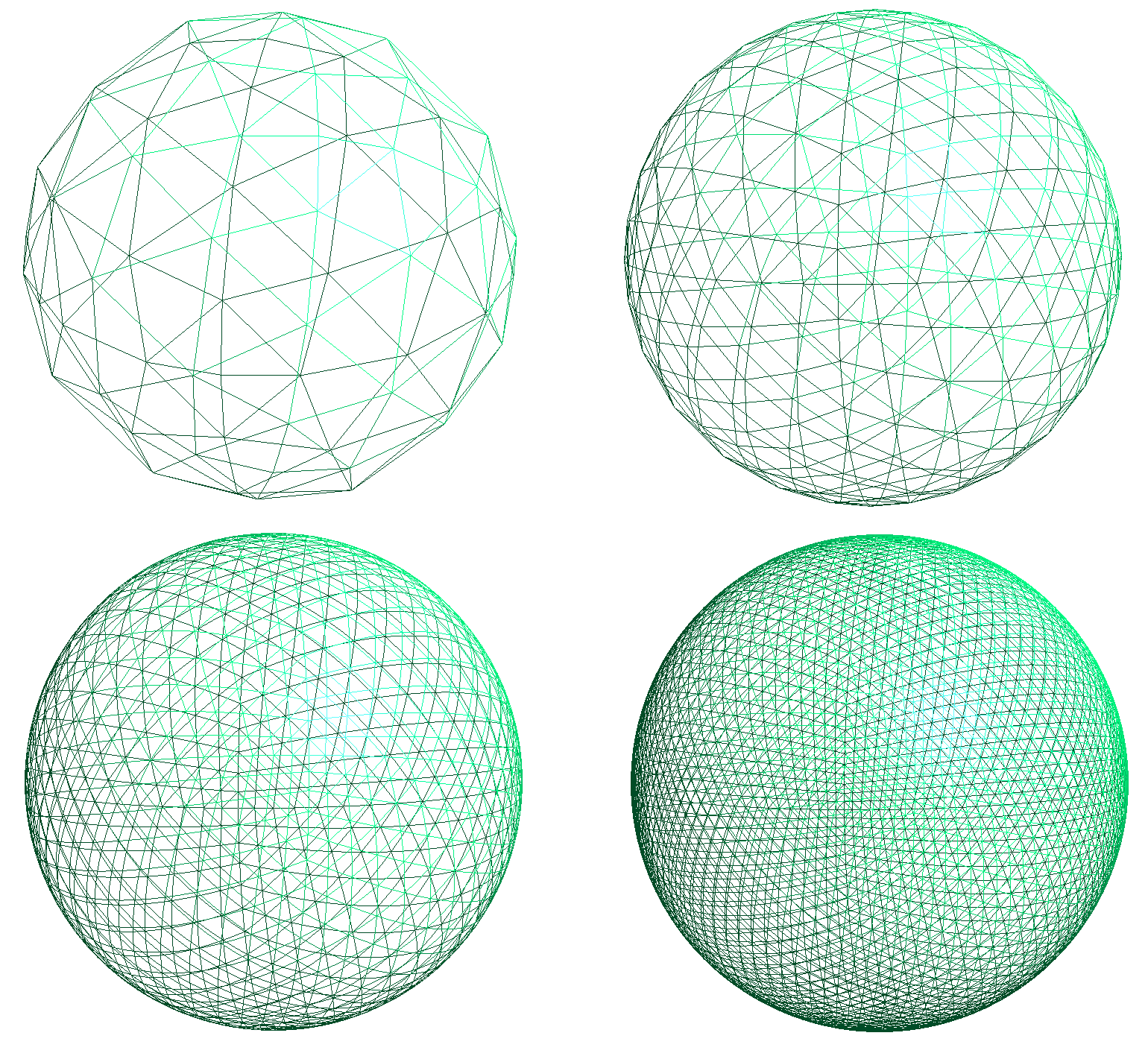}
\caption{Triangulation with (top left) $N=128$, (top right) $N=512$, (bottom left) $N=2048$, (bottom right) $N=8192$.}
\label{figure:mesh-all}
\end{center}
\end{figure}

For this example, we perform a convergence test for computing the two-point function between $x = (1,0,0), y=-x \in \bb{S}^2$ using piecewise linear, quadratic and cubic elements. The reference solution $r^*$ is computed using piecewise quartic elements with $h_N = 10^{-3}$, where $h_N$ is a proxy for the characteristic length $h$ on a sphere triangulated by $N$ elements,
$$ h \sim h_N := \sqrt{\frac{4\pi}{N}}. $$
For example, the triangulation for several values of $N$ are shown in \Cref{figure:mesh-all}.

\begin{figure}[H]
\begin{center}
\includegraphics[width=110mm]{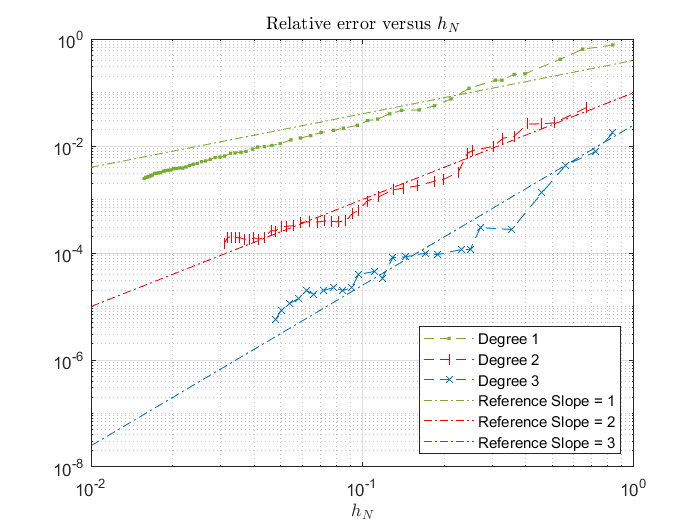}
\caption{Convergence test for piecewise linear, quadratic, and cubic elements used to obtain the two-point correlation function for a scalar field on the conformal two-sphere.} 
\label{figure:conv}
\end{center}
\end{figure}

The relative error is defined by
$$ \text{Relative error} = \frac{|\langle \phi(x)\phi(-x)\rangle - r^*| }{r^*} $$
where $\langle \phi(x)\phi(-x)\rangle$ is the computed value for the two-point function given by solving the above linear system and subsequently taking the dot product. The results of the test are shown in \Cref{figure:conv}. 

\begin{remark}
    In \Cref{figure:conv}, note that the piecewise linear elements appear to converge faster than first order. This phenomena, in which the observed convergence is faster than the expected convergence, is known as superconvergence \cite{super1994}. In particular, for a discussion of superconvergence on surfaces (including the sphere, as in this example), see \cite{super2010}.
\end{remark}

This is an example of how finite element techniques can be used to formulate lattice field theories on curved spaces, both in the possibility of introducing a Riemannian metric and also in allowing meshes on embedded manifolds inherently. One interesting direction to explore is examining how finite element techniques can be used to compute such correlation functions with nonlinear terms in the action without using statistical methods; in particular, linearization of nonlinear terms about ground states in the action may provide a path for computing perturbations to the correlation function about the linearization. This would involve successive linearization and subsequent solves of the linear system $(K+M)\vec{z} = \vec{b}$ discussed above.

\subsection{A remark on more general finite element spaces}\label{sec:general-fem}
To encompass a broader range of quantum field theories in the Galerkin discretization, we can relax two of the definitions made in \Cref{sec:galerkin}. First, one can relax the requirement that the discrete action $\bb{S}_h: X_h \rightarrow \bb{R}$ is the restriction of the continuum action to the finite-dimensional subspace, $S \circ i_h$, and instead allow it to be, more generally, some consistent $\mathcal{O}(h^p)$ approximation of $S$. This allows the flexibility of adding counter terms in the action such as in the Symanzik improvement program \cite{Improvement1,Improvement2}
as well as allowing quadrature approximations of integrals involving higher order nonlinear terms appearing in the action which is useful from a finite element perspective \cite{quad1, quad2, quad3}.

The subsequent definitions made in \Cref{sec:galerkin} follow similarly. For example, for such a choice $\bb{S}_h$, we define as before the discrete action on degrees of freedom 
$$ \mathfrak{S}_h = \bb{S}_h \circ \bb{E}_h^{-1} : \m{F}_h \rightarrow \mathbb{R}, $$
and the associated Galerkin partition function as before,
\begin{equation}\label{eq:path-integral-galerkin-discrete-general}
    Z_h = \int_{X_h} D\phi_h \exp(-\bb{S}_h[\phi_h]) = \int_{\m{F}_h} \prod_{i=1}^{\dim(X_h)} d\phi^i_h \exp(-\mathfrak{S}_h[\vec{\phi}_h]).
\end{equation}

Furthermore, in lattice gauge theories, group variables are used conventionally as opposed to Lie algebra variables \cite{Creutz23, He2012, Smit02}. From our perspective, this can be seen as working with functions of the degrees of freedom as opposed to the degrees of freedom themselves. Thus, we can more generally consider a measurable mapping $L: \m{F}_h \rightarrow Y_h$ from the space of degrees of freedom into a topological space $Y_h$. Supposing we have some consistent approximation $S_{Y_h}[y_h]$ of the action in terms of variables $y_h \in Y_h$, we define the partition function as
$$ Z_h = \int_{Y_h} dy_h \exp(-S_{Y_h}[y_h]) $$
where the measure $dy_h$ is defined as the pushforward measure of the measure $\prod_{i=1}^{\dim(X_h)} d\phi^i_h$ on $\m{F}_h$ by $L$. This additional flexibility in working with $Y_h$, obtained by a mapping from the degrees of freedom to $Y_h$, allows, for example, one to consider the Galerkin formulation in terms of group variables instead of Lie algebra variables as in standard formulation of lattice gauge theories, as we will see below.

\textbf{Lattice Gauge Theories.} Let us contextualize the more general definitions made above in the context of lattice gauge theories. For gauge field theories, the variables are $\mathfrak{g}-$valued $k-$forms on a domain $D$ (e.g., for quantum electrodynamics and quantum chromodynamics, $k=1$), denoted $A \in \Omega^k(D; \mathfrak{g})$, where $\mathfrak{g}$ is the Lie algebra of some compact Lie group $G$. Let $\mathcal{T}_h$ denote a simplical complex on the domain $D$ (or, alternatively, a cubical complex as is used in standard lattice gauge theory treatments) and let $\mathcal{T}^k_h$ denote the set of $k-$dimensional simplices (resp., $k-$faces) in $\mathcal{T}_h$.  Let $\Omega^k_h = W^k(\mathcal{T}_h; \mathfrak{g})$ denote the space of Whitney $k-$forms on $\mathcal{T}_h$. In this setting, the degrees of freedom can be viewed as $\mathfrak{g}-$valued variables at each $k-$simplex, given by integration of the field variable $A$ over such a $k-$dimensional simplex; particularly, this is given by the de Rham map. Denoting $N_k := |\mathcal{T}^k_h|$ as the number of $k-$simplices in $\mathcal{T}_h$, the degrees of freedom are
\begin{align*}
R^k : \Omega^k &\rightarrow \mathfrak{g}^{N_k}, \\
      A &\mapsto \left(\int_T A \right)_{T \in \mathcal{T}^k_h}.
\end{align*}
That is, elements of the space of degrees of freedom $\m{F}_h = \mathfrak{g}^{N_k}$ are given by associating a Lie algebra element in $\mathfrak{g}$ to each $k-$simplex in $\m{T}^k_h$. 
\begin{remark}
    Other geometric quantities appearing in a lattice gauge theory can be similarly expressed in the Galerkin framework as integration over simplices, possibly of different dimension. For example, the field strength tensor $F$, i.e., curvature form, in the continuum theory is defined as the exterior covariant derivative of $A$, i.e., $F = D_A A$. In the Galerkin formulation, this corresponds to a map
    \begin{align*}
    F^{k+1} : \Omega^k &\rightarrow \mathfrak{g}^{N_{k+1}}, \\
          A &\mapsto \left(\int_S D_A A \right)_{S \in \mathcal{T}^{k+1}_h},
    \end{align*}
    where $\mathcal{T}^{k+1}_h$ denotes the set of $(k+1)-$dimensional simplices in $\mathcal{T}_h$ and $N_{k+1} := |\mathcal{T}^{k+1}_h|$. In \cite{LUSCHERnpb}, an admissibility criterion was imposed on the field strength of the form $|F_{\mu\nu}(x)|<\epsilon$ for all $x,\mu,\nu$ in the setting of $U(1)$ gauge fields, i.e., $k=1$ and $G = \text{U}(1)$. This admissibility condition can be interpreted as a uniform bound on the field strength over all plaquettes. In the Galerkin framework, an analogous admissibility condition would be a bound 
    $$ \left|\int_S D_A A \right| < \epsilon \text{ for all } S \in \mathcal{T}^{2}_h,$$
    which, similar to \cite{LUSCHERnpb}, is a uniform bound on the field strength over all $2-$dimensional simplices.
\end{remark}

Conventionally, lattice gauge theories work with the exponential of these degrees of freedom, defined at each ``lattice site" of dimension $k$ (i.e., at each $k-$simplex),
$$ U_T := \exp\left(-\int_T A \right),\quad T \in \m{T}^k_h,$$
and use an action defined in terms of these ``link" variables $\{U_T\}_{T \in \mathcal{T}^k_h} \in G^{N_k}$, instead of the degrees of freedom, such as the Wilson action on cubical meshes \cite{Creutz23, He2012, Smit02} or higher order analogues obtained by Symanzik improvement \cite{Improvement1, Improvement2}. A simplicial analogue of the Wilson action is investigated in \cite{ChHa12} based on this first-order Whitney form approximation, particularly for $k=1$. As defined above, the measure for the partition function on $G^{N_k}$ is the pushforward of the measure on $\m{F}_h$; this becomes the $N_k-$fold product of the Haar measure on $G$, arising from the well-known fact that the pushforward of the standard measure on $\mathfrak{g}$ by the exponential map is the Haar measure on $G$. 

This shows how the standard lattice gauge theory formulation can be recovered from the Galerkin framework, using a first-order approximation space for $k-$forms given by Whitney forms. However, the Galerkin framework allows the possibility of higher-order approximation spaces. Such higher-order approximations of $k-$forms are well studied and an active area of research in the finite element literature, particularly in the context of finite element exterior calculus \cite{feec1,feec2,feec3}.
Such higher-order approximations are based upon weighting the integrals in the de Rham map
$$ A \mapsto \left(\int_T A w_i \right)_{T \in \mathcal{T}^k_h}, \quad i=1,\dots,s, $$
for an appropriate set of weighting functions $\{w_i\}_{i=1}^s$. This set of degrees of freedom would then give multiple link variables at each $k-$simplex corresponding to each weight,
$$ U_T^i := \exp\left(-\int_T A w_i \right), i=1,\dots,s, T \in \m{T}^k_h. $$
These can then be used to construct higher order versions of the Wilson action for cubical meshes or similarly, of the simplicial action introduced in \cite{ChHa12} based on Whitney forms. Although we will not go into further detail here, we leave this as an interesting research direction, in using the well-studied techniques of finite element exterior calculus to construct higher-order approximations of lattice gauge theories. The finite element exterior calculus framework may be particularly suitable to discretization of quantum field theories on topologically non-trivial spacetimes, due to their cohomology preserving properties \cite{feec1,feec2}. Furthermore, recent work on discrete vector bundles, their connections, and relations to finite elements could be insightful for such a study \cite{vec1,vec2,vec3}. 

A related direction is the Galerkin discretization of fermionic field theories. As mentioned in \Cref{sec:intro}, linear finite element discretizations of fermions were explored in \cite{QFE0, QFE1}. It would be interesting to develop higher-order approximations based on higher-order finite element spaces. The work of \cite{LeSt16}, in which the authors consider discretization of the Hodge--Dirac operator in the context of the finite element exterior calculus framework, may prove insightful towards this end.

\section{Conclusion}
In this work, we presented a mathematical framework for the Galerkin formulation of path integrals in lattice field theory. By taking the degrees of freedom associated to a Galerkin discretization as the fundamental field variables in the path integral, we show how many concepts in traditional lattice field theory can be viewed from a Galerkin perspective; this opens the possibility to using, e.g., finite element methods to construct higher-order approximations of path integrals and using the well-developed numerical techniques of the finite element method for lattice field theory calculations. Finally, we provide some possible future research directions.

As discussed in \Cref{sec:general-fem}, one possible direction is to utilize this framework to develop higher-order approximations of lattice gauge theories and fermionic theories using higher-order finite element approximation spaces. Another interesting direction to explore, which is particularly suited to finite element methods, is the use of multigrid and multilevel methods \cite{multigridMC1,multigridMC2,multigridMC3,multilevelMC} to develop efficient algorithms when paired with Markov Chain Monte Carlo methods for computing expectation values. Furthermore, by framing path integrals in terms of the Galerkin perspective, tools from functional analysis can be used to investigate convergence, consistency and related issues; we plan to perform such analysis in future work.

\section*{Acknowledgements}
The authors would like to thank the referees for their helpful comments and suggestions. BKT was supported by the Marc Kac Postdoctoral Fellowship at the Center for Nonlinear Studies at Los Alamos National Laboratory. BSS was supported by the DOE Office of Advanced Scientific Computing Research Applied Mathematics program through Contract No. 89233218CNA000001. Los Alamos National Laboratory Report LA-UR-24-32282.

\section*{Data Availability Statement}
The data generated in this study is available from the corresponding author upon reasonable request.

\bibliographystyle{plainnat}
\bibliography{lattice.bib}

\end{document}